# Tuning of Charge Order by Uniaxial Stress in a Cuprate Superconductor

## Authors

Laure Thomarat,[1,2] Frank Elson,[3] Elisabetta Nocerino,[4,5] Debarchan Das,[6] Oleh Ivashko,[7] Marek Bartkowiak,[1] Martin Månsson,[3] Yasmine Sassa,[8] Tadashi Adachi,[9] Martin v. Zimmermann,[7] Hubertus Luetkens,[10] Johan Chang,[11] Marc Janoschek,[1,11] Zurab Guguchia,[10]* Gediminas Simutis[1,8]*

## Affiliations

[1]Laboratory for Neutron and Muon Instrumentation, Paul Scherrer Institut, CH-5232 Villigen PSI, Switzerland
[2]École normale supérieure Paris-Saclay, 91190 Gif-sur-Yvette, France
[3]Department of Applied Physics, KTH Royal Institute of Technology, SE-106 91 Stockholm, Sweden
[4]Department of Materials and Environmental Chemistry, Stockholm University, SE-106 91 Stockholm, Sweden
[5]Laboratory for Neutron Scattering and Imaging, Paul Scherrer Institute, CH-5232 Villigen PSI, Switzerland
[6]Laboratory for Muon Spin Spectroscopy, Paul Scherrer Institute, CH-5232 Villigen PSI, Switzerland
[7]Deutsches Elektronen-Synchrotron DESY, Notkestraße 85, 22607 Hamburg, Germany
[8]Department of Physics, Chalmers University of Technology, SE-41296 Göteborg, Sweden
[9]Department of Engineering and Applied Sciences, Sophia University, Chiyoda, Tokyo 102-8554, Japan
[10]Laboratory for Muon Spin Spectroscopy, Paul Scherrer Institute, Villigen PSI, Switzerland
[11]Physik-Institut, Universität Zürich, Winterthurerstrasse 190, CH-8057 Zürich, Switzerland
*Corresponding authors. Email: zurab.guguchia@psi.ch, gediminas.simutis@psi.ch

## Abstract

Strongly correlated electron materials are often characterized by competition and interplay of multiple quantum states. For example, in high-temperature cuprate superconductors unconventional superconductivity, spin- and charge-density wave orders coexist. A key question is whether competing states coexist on the atomic scale or if they segregate into distinct "islands". Using X-ray diffraction, we investigate the competition between charge order and superconductivity in the archetypal cuprate $La_{2-x}Ba_xCuO_4$, around x = 1/8-doping, where uniaxial stress restores optimal 3D superconductivity at $\sigma_{3D} \approx 0.06$ GPa. We find that the charge order peaks and the correlation length along the stripe are strongly reduced up to $\sigma_{3D}$, above which they stay constant. Simultaneously, the charge order onset temperature only shows a modest decrease. Our findings suggest that optimal 3D superconductivity is not linked to the absence of charge stripes but instead requires their arrangement into smaller "islands". Our results provide insight into the length scales over which the interplay between superconductivity and charge order takes place.

**Introduction**

An intriguing feature of high temperature cuprate superconductors is the competition and cooperation between stripe order and superconductivity. Stripe order refers to a pattern of charge and spin density modulations that emerge at low temperatures in these materials (1–5). This ordering phenomenon has substantial impact on the properties of the cuprates. Yet, the exact nature of the stripe order and its relationship to the underlying electronic structure is still the subject of intense debate. There is strong evidence that stripe order competes with three dimensional superconductivity in the cuprates, leading to a suppression of the critical temperature for superconductivity (6, 7). However, there are also indications that the stripe order may play a more subtle role in the superconducting state, such as stabilizing certain exotic types of superconducting pairing (8, 9). Understanding the interplay between stripe order and superconductivity in the cuprates is thus a crucial step towards a comprehensive understanding of these fascinating materials.

As the different ground states in these systems are subtly balanced, small external perturbation can be used to select between the emergent phases. Recently, uniaxial stress has emerged as an effective surgical tool to manipulate these systems. In a striking example, a three-dimensional charge density wave was observed in $YBa_2Cu_3O_{7-x}$ at zero magnetic field (10, 11). On the other hand, in $La_{2-x}Sr_xCuO_4$ the population of different stripe order domains could be achieved to unequivocally demonstrate the uniaxial nature of the stripe order and exclude any reasonable checkerboard models (12–14).

The best way to understand the complex interplay between these competing states is to study a system, where the lattice, charge and spin degrees of freedom as well as superconductivity has well-defined states that clearly change in the phase diagram. As such, $La_{2-x}Ba_xCuO_4$ is an excellent test system - a dramatic suppression of 3D superconducting transition temperature $T_{SC}$ is observed around 1/8 doping, where the charge and spin order is most pronounced as seen in the phase diagram of Fig. 1**a** (15-18). In this range, however, superconductivity persist in the two-dimensional layers of copper-oxygen planes (19). At the same doping levels, the system undergoes a structural phase transition towards a low-temperature tetragonal (LTT) structure, where the orientation of copper-oxygen octahedra alternates between adjacent layers (20). Such a restructuring of the lattice potentially locks in the stripe order, consequently frustrating the Josephson coupling between the individual superconducting layers (3, 21).

Thus, the presence of multiple structural phases across the temperature and doping range raises the question of what are the underlying driving phases. On one hand it has been suggested that the LTT phase is a host structure for the stripe formation (22). However, charge order has been repeatedly observed outside this structural phase throughout various cuprate systems (23–31). As the stripe order has been shown to be incompatible with the orthorhombic phase (32), some other deviations from the pure orthorhombic symmetry must be present in the system. The establishment of the Low-Temperature Less-Orthorhombic (LTLO) phase was therefore suggested to take place in $La_{2-x}Ba_xCuO_4$ (15). Here, the tilt axis of the copper-oxygen octahedra points along an intermediate in-plane direction (33).

The utility of $La_{2-x}Ba_xCuO_4$ as a model system has been further proven by the recent observation that 3D superconducting transition temperature is vastly increased upon moderate uniaxial compression of the superconducting planes along the copper-copper direction (17), as seen in Fig. 1**b**. Similarly, uniaxial-stress induced enhancement of $T_C$ and effect on charge order was also reported in a Nd-doped $La_{2-x}Sr_xCuO_4$ (34-36). Concomitant with the increase of superconductivity, a suppression of the magnetically ordered volume fraction in $La_{2-x}Ba_xCuO_4$ has been observed by both muon spin rotation (16, 17) as well as neutron scattering (37). A recent structural study has revealed that the applied stress also suppresses the tetragonal structure and lifts the geometric frustration (16). The remaining question is: what happens to the charge order upon this compression? In this study, we use hard X-ray diffraction coupled to in-situ uniaxial stress tuning (38) to directly determine the effect on the charge order as illustrated in Fig. 2**a**. This allows us to complete the phase diagram of $La_{2-x}Ba_xCuO_4$ under uniaxial stress and describe the evolution of its structural, charge and magnetic phases.

We find that the intensity of the charge order peaks as well as the correlation length along the stripe direction is strongly reduced with uniaxial stress up to the critical value $\sigma_{3D}$, defined as the pressure at which the LTT phase is completely suppressed, above which they stay constant. On the other hand, the transition temperature of the charge order shows only a modest decrease upon approaching $\sigma_{3D}$, above which it also stays unchanged. Our findings reveal that establishment of optimal 3D superconductivity does not require a full suppression of charge-stripes but rather weakening them to an optimal limit. Our findings point toward a remarkable cooperation between superconducting pairing/coherence and spin/charge/lattice degrees of freedom.

**Results**

A key characteristic of charge stripe order is the emergence of diffraction peaks at $\mathbf{Q}_{CDW} = \boldsymbol{\tau} \pm \mathbf{q}_{CDW}$ with $\boldsymbol{\tau}$ being fundamental Bragg reflections and $q_{CDW} = (\delta,0,1/2)$ and $(0,\delta, 1/2)$ describing the charge order incommensurability, which depends on doping (15). The structure factor implies that the charge order peak amplitudes vary in strength across different Brillouin zones (39). Due to our experimental setup that includes a cryostat and a magnet with only a small aluminum windows for the X-ray path, the background scattering also varies with momentum. Generally, contaminating powder rings originating from the sample environment are less pronounced in zones with large momenta ($|Q| > 15$ r.l.u.). At ambient stress, we have inspected multiple charge order peaks and identified zones with the best intensity to background ratio. Therefore, we concentrate on a Brillouin zone defined by $\boldsymbol{\tau} = (2,2,17)$ with respect to the high-temperature tetragonal notation (a = b = 3.78 Å). Additionally, for cross checks, a second peak in the (1, 1, 9) Brillouin zone was measured for consistency (40).

Our main results are summarized in Fig. 2**b**, where the 3-dimensional reciprocal scattering volume is projected onto the (2,k,l) plane with an integration of 0.005 r.l.u. perpendicular to the plane. First, we note that in all of the measurements a charge order peak can be identified and is gradually suppressed with strain. For a clearer illustration of the suppression, we also present the one-dimensional cuts through the peak in Fig. 3**a**. Analogous measurements under magnetic field are presented in Fig. 3**b**. The data in these plots is obtained by integrating the intensity around the peak, perpendicular to the cut by 0.005 r.l.u. in h direction and 0.05 r.l.u. in l direction. We further note that due to the setup of our experiment, we likely do not have the resolution to determine possible subtle shifts of the peak position, but we can identify the upper limit of the shift as 0.03 r.l.u. for the h position, 0.01 r.l.u for the k position and 0.1 r.l.u for the l position. For the purpose of presentation, we therefore fix the positions to the ambient stress value.

When we consider both the zero-field and applied-field cases, the peak amplitude is reduced as already seen in the 2D-cuts and the width of the peak increases as the stress is applied. In order to perform a more detailed study of this trend, we have fit the data to a three-dimensional model as well as in several different limited cuts, obtaining a consistent picture, independent on the analysis methods.

The obtained fit results are presented in Fig. 4. At ambient stress, the peak intensity increases with decreasing temperature, consistent with earlier measurements (15). As the stress is increased, the onset temperature of the charge order decreases, as shown in Fig. 4**a**. At ambient stress, the charge order is found to arise at temperatures that coincide with the onset of the structural transition towards the tetragonal symmetry (15, 16). Remarkably, even though the full low-temperature tetragonal structure is suppressed with a modest critical stress of $\sigma_{3D} \approx$ 0.06 GPa (16), the charge order persists up to the highest applied stress, with the change saturating above $\sigma_{3D}$. We note, that the same behavior qualitatively is observed for spin order (17). Inspecting the structure evolution in detail, we observe that at elevated stress value, the charge order tracks the onset of the LTLO structure, where the orthorhombicity is reduced (Fig. 1**b**). Moreover, as stress is increased, the intensity of the scattering saturates at lower values than in the unperturbed case. This is explicitly shown in Fig. 4**b**, where it becomes clear that the fast suppression of the intensities saturates beyond $\sigma_{3D}$ and the scattering with a finite value persists up to the highest measurable stress. In addition to the amplitude of the peaks, we also investigate the widths of the peak in all three directions in the reciprocal space as the stress is increased. This is shown in Fig. 4**c** and reveals a well-defined trend - the correlation length (defined as the inverse of HWHM) along the stripe propagation direction decreases with the increase of the stress, whereas the correlation lengths of the charge order in the other two directions remain unchanged.

**Discussion**

Superconductivity in $La_{1.885}Ba_{0.115}CuO_4$ is enhanced with uniaxial stress (16, 17) as the magnetic and charge order is reduced. This reduction does not result in a complete suppression however, the magnetic and charge modulation phases are persistently present in the system throughout the studied stress range.

We first address the differences between the previously studied spin channel (16) and the presently measured charge channel of the stripe order. In contrast with the recovery of the volume fraction of the spin order reported previously (16) towards low temperatures, we find that the intensity of the charge order peaks saturates and stays constant to the lowest measured temperatures. This suggests that the superconductivity and charge order manifest effectively in the same stripe, whereas the spin order concerns a larger volume of space.

We now turn towards understanding the mechanism that increases the 3D d-wave superconductivity upon the uniaxial compression of the system. A plausible mechanism is the reduction of the frustration of the Josephson coupling (8) when we pressurize and suppress the LTT phase (16). However, a careful examination of the relevant temperatures for structural distortions and the emergence of the charge order points to a different scenario. While the full formation of the LTT phase is suppressed with moderate stress of $\sigma_{3D} \approx 0.06$ GPa, the charge order persists up to the highest measured values, at least three times of $\sigma_{3D}$. Moreover, we find that the onset of the charge order tracks the onset of the LTLO structure. As such, this observation indicates that the LTT phase is not a requirement for the formation of stripe order. However, it enhances the macroscopic extent of the stripe order islands. Such a conclusion is in line with the persistence of finite-volume stripe order phases in Sr-doped cuprate systems (12, 41, 42), where the LTT phase is not present. Yet, it has been recently pointed out that the stripe order symmetry is incompatible with the low-temperature orthorhombic crystal structure (32). It is therefore conceivable that the orthorhombic structure is accompanied by weak monoclinic distortions that enable the formation of the stripe order. As local distortions can be averaged over the whole sample, it is possible that they are "disguised" as an average LTLO structure.

This leads us towards answering the main question: what is the mechanism behind the competition between the superconductivity and the charge-spin stripes? One scenario would be that the stripes align between the adjacent layers. This would shift the peak intensity away from the half-integer towards integer values, perpendicular to the cuprate planes. We, however, observe that there is no change in the position or the width of the CDW peak along $c$-direction, even though the LTT phase is completely suppressed at this value of stress (16). Similarly, charge order persists also when the four-fold rotational symmetry is recovered upon application of high hydrostatic pressure (43). The effect may be much more subtle as the charge stripes can be pinned by the domains walls between different orthorhombic twin domains (44) and can even be impacted by the structural properties at elevated temperatures (45). Therefore, we conclude that the key element that allows the increase of the superconducting temperature and formation of coherent $d$-wave superconductivity is the optimization of the sample volume where the stripe order is unfavorable.

This is further corroborated by our measurements investigating the role of magnetic field. In the doping regime where the superconductivity is favorable compared to the pinned stripe order, the magnetic field can greatly increase the intensity of stripe order peaks (46). In $La_{1.885}Ba_{0.115}CuO_4$, the magnetic field has very little effect as the stripes are already fully formed. In the "competition-only" scenario, the application of magnetic field could recover the stripes after they are reduced due to uniaxial stress. We find that the magnetic field has no effect on the stripe order in the uniaxially pressurized phase. This observation implies that likely the stripe order and superconductivity achieve an energetically favorable balanced state upon the application of uniaxial stress and therefore the field effect is absent. Our conclusion is also consistent with, and a natural co-explanation of the recent measurements, where proton irradiation was used to induce disorder in $La_{1.875}Ba_{0.125}CuO_4$ crystals (47). The irradiation simultaneously increased the superconducting transition temperature and decreased the correlation length of the stripe order, pointing to the importance of the optimal length scale upon which the stripes have to be correlated for the maximization of coherent 3D superconductivity in cuprates. Similar conclusion has been reached in a recent study where an inverse effect has been reported – the enhancement of charge order through quenching of superconductivity in $YBa_2Cu_3O_{6+x}$ (48). Finally, very recent RIXS measurements have suggested that among the multiple intertwined orders in cuprates, pair density wave (PDW) exhibits a separate subharmonic order at temperatures above the long-range superconductivity (49). In the future, it would be revealing to study the evolution of such signatures using uniaxial pressure methods and compare with the changes in stripe order.

In summary, we have directly shown that the increased 3D-superconductivity temperature in $La_{1.885}Ba_{0.115}CuO_4$ is accompanied by the suppression of the charge ordered phase, which nevertheless persists in a finite volume of the sample. The state obtained above the critical pressure is very stable to further compression as well as magnetic fields and suggests a subtle cooperation between the intertwined orders. This points towards the need to include real-space volume inhomogeneity in the theoretical description of emergent phases in cuprates.

**Materials and Methods**

Following the method devised in Ref. 50, polycrystalline samples of $La_{2-x}Ba_xCuO_4$ with $x = 0.115$ were synthesized using the conventional solid-state reaction method, which involved the use of $La_2O_3$, $BaCO_3$, and $CuO$ as starting materials. The powders obtained were then subjected to scintering and verified to be single-phase through powder X-ray diffraction analysis. The single crystals of $La_{2-x}Ba_xCuO_4$ with $x = 0.115$ were grown using the traveling solvent floating-zone method. In this technique, a small molten zone is created between two counter-rotating rods, which melt the material and then cool it down to form a single crystal. The crystal was then aligned and cut into a cuboid shape with the tetragonal [0,0,1] and [1,1,0] directions spanning the scattering plane. Stress was applied to the crystal at an angle of 45° to the Cu-O bond direction.

X-ray experiments were performed on P21.1 beamline at Petra-III synchrotron radiation source in DESY. Photon energy of 101626 eV (corresponding to 0.122 Å) was used to penetrate the complex sample environment that allowed us to cool down to 4.5 K and apply horizontal field up to 10 T. The sample was pressurized using a recently developed in-situ uniaxial device, based on linear motor actuator (32). The constant force on the sample was maintained using a separate feedback system and the temperature was measured independently on the uniaxial cell next to the sample. As the magnet has limited window (opening of 10° in four perpendicular directions of the scattering plane), the sample was fixed in an orientation that allowed us to reach several most intense charge order peaks. This meant that the center of the window was oriented at 30° with respect to the crystal **c** axis. We have applied a magnetic field of 10 T to the sample, which is the value quoted in Fig. 3**b**, but it is important to note that the magnet is designed in a way, where the field can only be applied perpendicular to the windows that allow the X-rays to get through. Taking into account the previously mentioned rotation of the sample meant That the effective field along the **c** axis was therefore B = 8.7 T. Further, as the magnet contributes substantial background, we have performed additional measurements at elevated temperatures where the charge order is no longer present and used that as a background, which we subtracted to obtain the data to be analyzed.

**Acknowledgments**

We thank N. B. Christensen and D. G. Mazzone for valuable discussions about cuprates under pressure and thank Philipp Glaevecke and Olof Gutowski for their assistance with the technical aspects of the experiment.

**Funding:**

GS was partially supported through funding from the European Union's Horizon 2020 research and innovation programme under the Marie Sklodowska-Curie grant agreement No 884104 (PSIFELLOW-III-3i). Y.S. is funded by the Swedish Research Council (VR) through a Starting Grant (Dnr. 2017-05078). Y.S. and G.S. further acknowledge funding from the Area of Advance-Material Sciences from Chalmers University of Technology. E.N. is supported by the Swedish Foundation for Strategic Research (SSF) through the SwedNess grant SNP21-0004. J.C. would like to thank the Swiss National Science Foundation for the financial support.

**Figures**

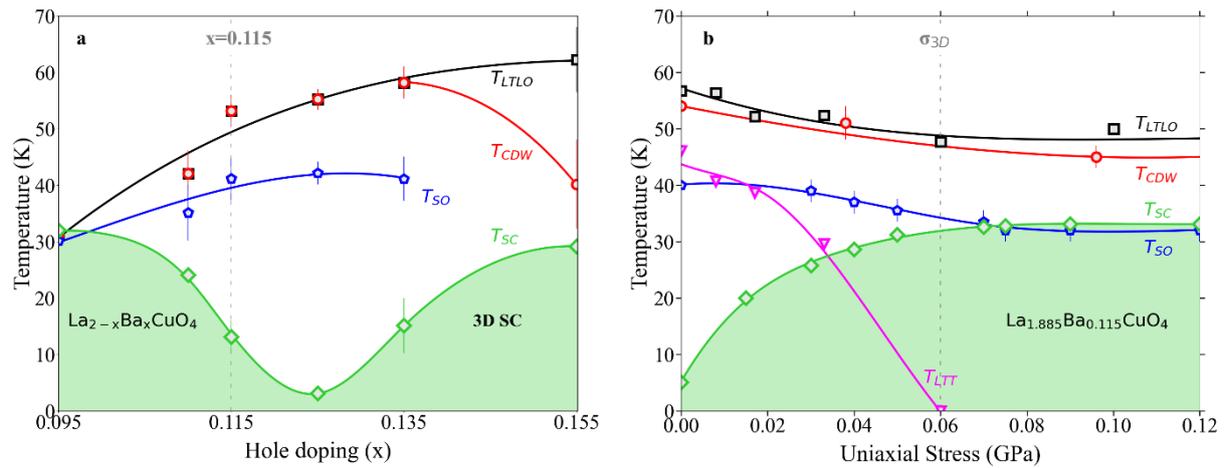

**Fig. 1. Phase Diagram of La$_{2-x}$Ba$_x$CuO$_4$.** **(a)** Temperature versus hole-doping phase diagram from (15), showing the multiple electronic and structural phases and indicating with a grey dashed line the newly-explored part of the phase diagram, accessed by uniaxial tuning. **(b)** The complete phase diagram as a function of uniaxial stress, around the doping of x = 0.115, where the multiple degrees of freedom and corresponding phases are modified. As superconducting temperature is increased with stress, the transition temperatures to Charge Density Wave (CDW) and the spin order (SO) are suppressed, together with the reduction of macroscopically occupied volume fraction as discussed in the text. Notably, the structural LTT phase is completely suppressed at moderate stress, yet the electronic stripe order persists throughout the studied pressure range, albeit with a reduced volume fraction. All changes saturate above the meager stress of $\sigma_{3D} \approx 0.06$ GPa. The transition temperatures for the superconductivity, spin order and structural transformations for panel **(b)** are taken from (16), where exactly the same sample was used.

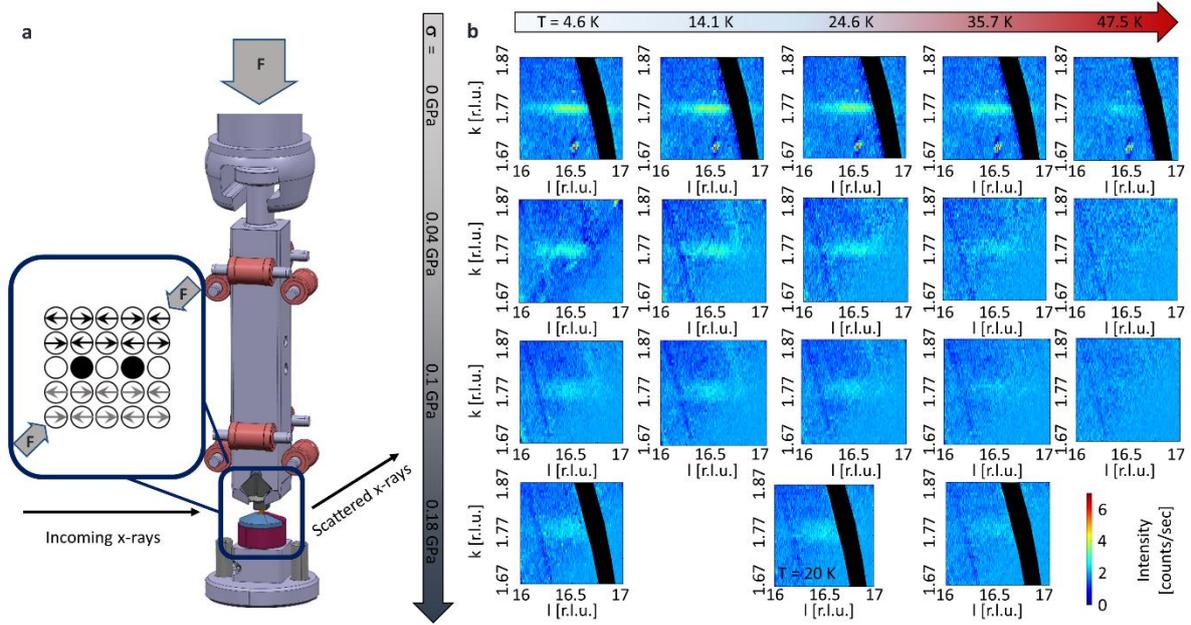

**Fig. 2. Charge Order Under Uniaxial Stress. (a)** The experimental setup used to measure the charge order peaks in $La_{1.885}Ba_{0.115}CuO_4$. The incoming X-rays that hit the uniaxially pressurized sample and the resulting diffraction pattern is collected by a two-dimensional detector, which allows a reconstruction to intensity as a function of the position in the reciprocal space. The close-up of the sample region illustrates that stress is applied along copper-copper direction, diagonally to the stripe propagation direction. **(b)** Development of the CDW as a function of temperature (horizontal axis) and uniaxial stress (vertical axis). Elastic scattering arising from the charge order manifests as a cigar-shaped peak in the reciprocal space. The two-dimensional reciprocal-space maps show how the intensity of the scattering decreases as a function of temperature at all stress values. Further, the intensity of the charge order peak decreases upon the increase of uniaxial stress. The black arcs and a sharp spot at k = 1.7 and l = 16.5 are due to background arising from sample environment.

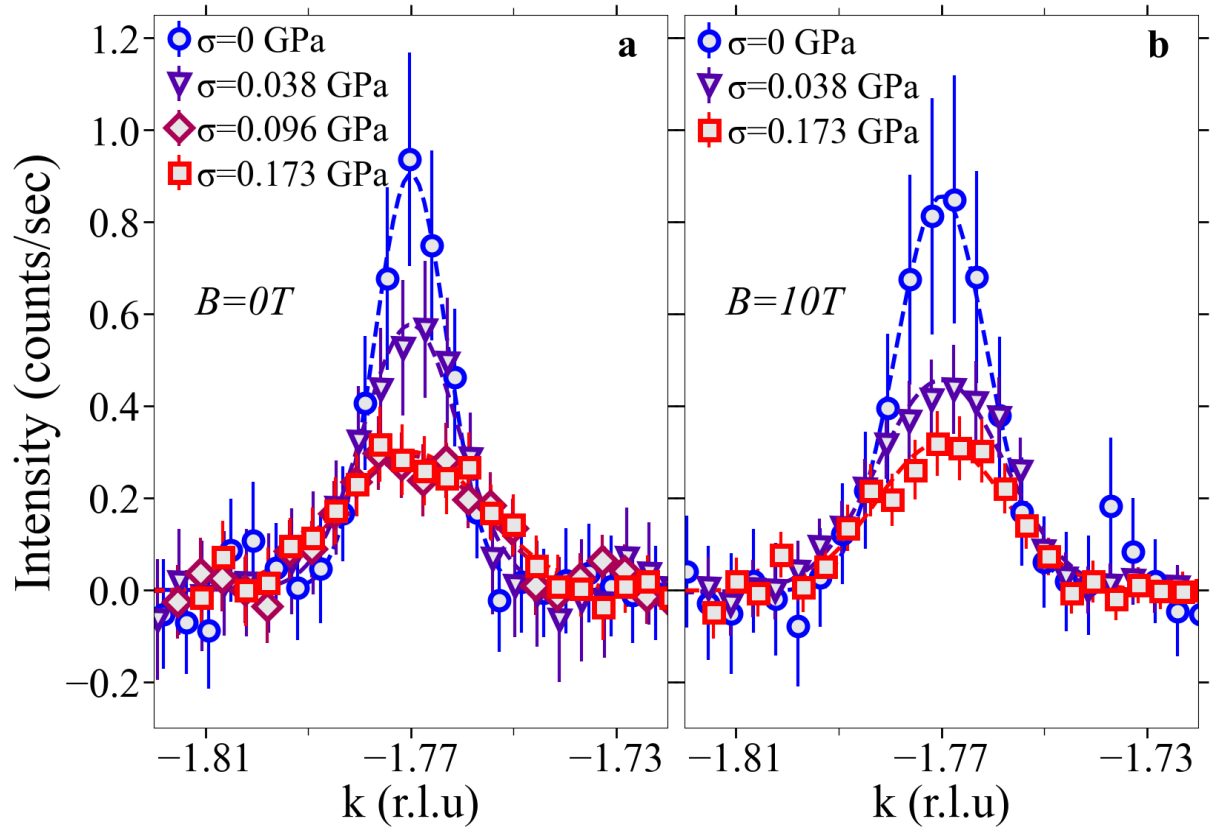

**Fig. 3. Tunability of charge order with uniaxial stress.** A one dimensional cut at base T = 4.6 K through the charge order peak at zero (**a**) and applied magnetic field (**b**) at various values of applied stress after subtracting the background. In both cases, the peak amplitude is drastically reduced and the widths of the peaks increase, corresponding to the reduction of the correlation length in the propagation direction.

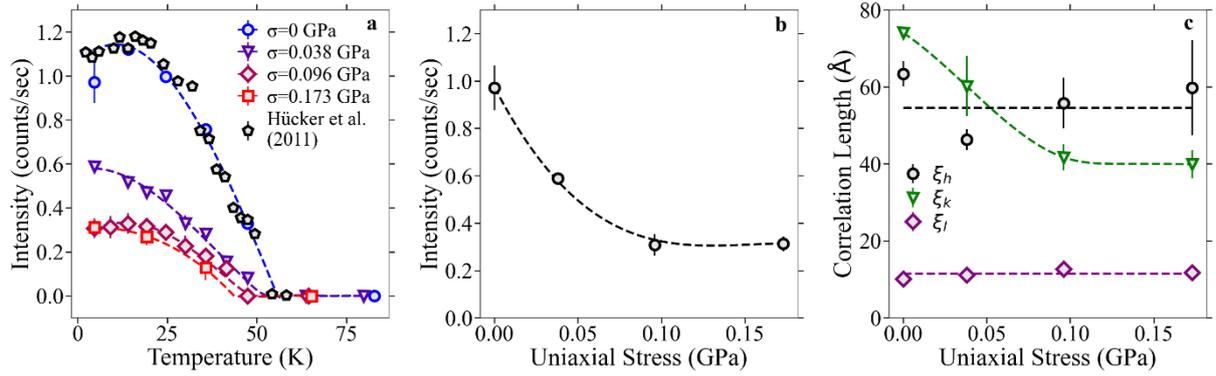

**Fig. 4. The evolution of charge-order parameters as obtained from the (2, 1.77, 16.5) peak.** (**a**) shows the amplitude of the peak as a function of temperature for the various values of applied stress. As a comparison, we also show the temperature dependence from **Ref.** (15), which is normalized to overlap with our zero-stress data. (**b**) shows how the base temperature peak amplitude varies as a function of applied stress. The stress-dependence of the correlation lengths for the different directions is shown in (**c**). The lines in all the panels are guides to the eye.